\documentclass[11pt,a4paper]{article}
\usepackage{amsfonts}
\usepackage{amssymb}
\usepackage{comment}
\usepackage{epsfig}
\usepackage{graphicx}
\usepackage{amsmath}
\usepackage{amsxtra}

\newcommand{\beq}{\begin{equation}}
\newcommand{\eeq}{\end{equation}}
\def\be#1\ee{\begin{align}#1\end{align}}
\newcommand{\ov }{\over }

\def\one{{\hbox{1\kern-.8mm l}}}
\def\vk{\vec{k}}
\def\vx{\vec{x}}

\title{Eliminating infrared divergences in an inflationary cosmology}

\vspace{2cm}
\author{{\Large Diego Chialva$^1$ and Anupam Mazumdar$^{2, 3}$} \\
       {\em \normalsize $^1$Universit\'e de Mons, Service de Mecanique et gravitation, Place
       du parc 20, 7000 Mons, \hfill} \\
       {\em \normalsize Belgium \hfill} \\
       {\em \normalsize $^2$Physics Department, Lancaster University, Lancaster, LA1
       4YB, United Kingdom \hfill} \\
       {\em \normalsize $^3$Niels Bohr Institute, Copenhagen University,
         Blegdamsvej-17, Copenhagen-2100, Denmark \hfill} \\}
\date{}

\begin{document}




\maketitle

\thispagestyle{empty}

\begin{abstract}
We study the infrared divergences arising from gravitational loops in the standard cosmological perturbation theory. 
We provide a simple solution to the problem at all orders of cosmological perturbation theory by
redefining the perturbation theory in terms of a local observer. We propose to reformulate 
the standard perturbations in the {\it in-in formalism},  and obtain an infrared safe perturbation theory.
Our results do not depend on any infrared cutoffs or similar
parameters. We then present an explicit 
example of graviton one-loop corrections, and briefly discuss non-Gaussianities.
\end{abstract}

\newpage


\setcounter{page}{1}

\tableofcontents
 
\section{Introduction}

The primordial inflation~\cite{Guth:1980zm} is perhaps one of the most important
paradigms of the early Universe, for a recent review,
see~\cite{Mazumdar:2010sa}. Inflation    
is responsible for stretching the initial perturbations to the
observable scales in 
the cosmic microwave background radiation (CMBR), and seeding the
initial perturbations for the large scale structure
formation~\cite{WMAP}. Since the future observational constraints will 
provide a better understanding of the inflationary dynamics and its
potential, it is then important to reach 
the desired level of accuracy by studying higher order quantum
corrections to the  
cosmological perturbations. For a review on cosmological perturbation
theory, see~\cite{Mukhanov:1990me}. 

The correlators at loop level in an inflationary setup appear to be
plagued by infrared (IR) divergences for {\it soft} quantum 
modes whose wavelengths are extremely large, known as the super Hubble
fluctuations. The divergences appear when these fluctuations are 
summed over in the loops \cite{IRLiterature, Weinberg:2005vy, Seery:2010kh, Riotto:2008mv,
Dimastrogiovanni:2008af, Burgess:2009bs, Giddings:2010nc, 
Burgess:2010dd, Koivisto:2010pj}. Regularizing the
integrals via  IR cutoffs does not solve the problem, but turns 
the divergences into large {\it Logarithmic} corrections depending on
the cutoff (``box size''). There has been a debate about the question
if such correction are physical or not~~\footnote{ There are attempts to address the IR issue 
by studying the pre inflationary phase in the early Universe, which
modifies the long wavelength 
behavior of the solutions of the field equations, thus ameliorating or
even canceling  the IR divergences
\cite{Koivisto:2010pj,Marozzi:2011da}. However, 
this approach has a certain degree of arbitrariness, depending on the
choice of a specific pre-de Sitter scenario. For example, a possible
solution is to 
evolve the perturbations from a contracting phase to the expanding
phase in a singularity free bouncing cosmology, where gravity becomes
asymptotically free in the ultra violet regime~\cite{Biswas:2005qr}.  
Modification of the perturbation equations capable of ameliorating
the IR issue (as they affect the spectral index) can arise also
in models such as chain inflation, due to the interactions among the different
components of the system
\cite{Burgess:2005sb,Chialva:2008zw}.}.  

Typically, the IR divergences are a signal of an ill-posed physical
question, and therefore their 
resolution depends on our understanding of the physical system and the
approach we adopt to tackle the problem. For instance, an unphysical  
element in the approach which leads to the divergence could be due to
an erroneous definition of an initial and final vacuum in a scattering
process, as it happens in the case of soft photons or
gluons emitted as a result of any quantum
process~\cite{Weinberg:1965nx}. The IR divergences  
may also arise if the perturbation theory has been wrongly organized
without taking into account the relevant scales that make some contributions 
unsuppressed, as for the case of field theory at finite temperatures~\cite{LeBellac}. 

Another point of view on this problem  has instead considered
the dependence  on the box size (IR cutoff) of the
correlators as a physical input. The question then arises -- whether the 
correlators, depending on the box size, had then to be averaged over
a distribution of boxes partitioning the Universe on super Hubble scales, 
or if the correct physical interpretation would be to fix the box size to the desired 
scale of observation and keep the
Log-enhancements~\footnote{\label{criticdecompcutoff} This also
  seems to be the point in Ref.~\cite{Gerstenlauer:2011ti, Byrnes:2010yc}, as the
  authors distinguish between IR and non-IR perturbations with respect to some
  typical observer's scales $q_0$ and $L^{-1}$, and therefore the spectrum which they
  claim to be IR-safe would  then depend on this cut-off procedure. We
  comment more on this in section \ref{amendingprocedure}.}.     
Both these approaches have negative aspects -- the first one does not take
account of the fact that the observer is not capable of averaging
over boxes larger than its Hubble patch, while the second approach still
suffers from the ambiguity on how to define the box size (for example,
via comoving coordinates or physical ones), which therefore makes the
result for the correlators ambiguous.

In this paper, we will point out that the issue of IR divergences in
gravitational loops in the
cosmological correlators can be solved  
by defining a {\it local observer} who is responsible for measuring
the observable quantities. We 
will argue  that once the observer and  
observables are well defined, the IR divergences will turn out to be
an artifact of some unphysical assumptions, 
usually taken in the definition of the perturbation theory. 

Our implementation of the principle of locality will be different from
what discussed in other studies in the literature~\footnote{The issue
  of locality has been discussed, with a different approach, also in
  \cite{Urakawa:2009my, Urakawa:2010kr}. Their proposal, however,
  entails considering non-standard gauge transformations that are
  singular at large scales and results in
  new gauge conditions \cite{Urakawa:2009my} or in the necessity of
  using perturbation fields {\em different} than the usual curvature 
and tensor ones, with a prescribed set of boundary conditions,
in order to avoid the IR divergences \cite{Urakawa:2010kr}.}.
In fact, our resolution is to redefine the standard perturbation theory in
the in-in formalism  based on one physical principle:
\begin{itemize}
 \item {\it any observable quantity should be defined in terms of a local observer
  who is measuring those observables}.
\end{itemize}
As a result, the observable quantities are shown not to have IR divergences, 
nor to depend on the IR cutoffs. 

The paper is organized as follows; in section \ref{Ininformalism},
we start with a brief introduction of the standard in-in formalism and the
quantization of the perturbations. We then provide a general discussion of the
IR behavior of loop correlators in section \ref{GeneralBehaviour},
using the example of self-energy diagrams, and
discuss the ambiguities in the cutoff procedure for regularizing the integrals. 
We present our resolution of the IR issue in section
\ref{physicalproposal}. First, in section \ref{conceptualpoint}, we
discuss the nature of the true observables and how the in-in
formalism, as it stands, fails to account for them. Then, in
\ref{IRallorders} we show
that all IR divergences at all orders arise only from certain specific
contributions to the correlators, and can be traced back to the
definition of the correlators using the background unperturbed metric;
finally we discuss how to eliminate the IR divergences by 
amending the standard in-in computations in section
\ref{amendingprocedure}, and provide a detailed 
example at one-loop for a  specific case in section
\ref{oneloopexample}. Finally, we will briefly discuss  
the non-Gaussianities in section \ref{Npointfunctions} and conclude in  
section \ref{conclusion}.

\section{Basic formalism}\label{Ininformalism}

The appearance of large IR corrections in the 
perturbation theory in an inflationary background is directly linked to
the presence of super-Hubble fluctuations of light
fields~\cite{Linde:1983gd,Linde:1986fd,Vilenkin:1982wt,Vilenkin:1983xq,Linde:1993xx,Linde:2005ht},  
it is therefore not related to any specific property of the field,
i.e. graviton or scalar field, running 
inside the loops. Here, we will briefly introduce the basics of
perturbation theory as it 
is usually formulated using the in-in formalism and the path integral
quantization, for a review, see~\cite{Weinberg:2005vy}; at the same time, we
list our conventions. 

The cosmological perturbation theory begins with the 
definition of a threading and a slicing of a spacetime (foliation) in
the unperturbed background metric,  
which is fully homogeneous and isotropic. Then
all the fields (including the metric) are written distinguishing the background
and the perturbations according to the chosen threading and slicing,
for a review, see~\cite{Weinberg:2005vy}. 
The background is considered to be classical, while the perturbation
fields are quantized.

Let us consider a scalar field $\Phi$ (similar procedure will follow for tensor fields, 
gravitons, once a polarization tensor is specified), and write it in
terms of  the background $\phi_0(t)$ and the 
perturbations $\phi(\vec x, t)$ as:
 \beq \label{Backgroundandperturbations}
  \Phi(\vec x, t) = \phi_0(t) + \phi(\vec x, t) \, .
 \eeq
The field $\phi(\vec x, t)$ is expanded on a basis of eigenfunctions
$Y_{\vec k}(\vec x)$ of the
Laplace-Beltrami operator $\nabla^2$ with eigenvalues $-k^2$. Our conventions
are: $k$ is the comoving wavenumber related to the physical
momentum $p$ by 
 \begin{equation}\label{co-phys}
  k=ap\,,~~
 \end{equation}
and the conformal time is defined via the background metric 
 \beq
  ds^2= -dt^2+a^2\delta_{ij}dx^i dx^j.
 \eeq 
as
 \be
  d\eta =a^{-1} \, dt\,,
 \ee
 where $a$ is the scale factor. We set the reduced Planck mass,
 $M_\text{P}^2=1$. The field can then be promoted to a quantum
 field, written as~\footnote{We will denote quantum fields, both fundamental and 
composite, with $\widehat{~~}$\,.}:
 \beq \label{fieldexpansion}
  \widehat \phi(\vx, \eta) = \int {d^3k \ov (2\pi)^{{3 \ov 2}}} \left[Y_{\vk}(\vx) u_k(\eta)
    \widehat a_{\vk} +Y_{\vk}^*(\vx) u_k^*(\eta) \widehat a_{\vk}^\dagger \right]\,,
 \eeq
and quantized in the usual way,
 \beq
  [\widehat a_{\vk}^\dagger,~ \widehat a_{\vk'}] = \delta(\vec k-\vec k') \, .
 \eeq
The Whightman function is defined as:
 \beq \label{WhightmanfunctionModeexpansionState}
  W(\vx, \eta, \vx', \eta') = \langle \Omega| \widehat\phi(\vx, \eta)\widehat\phi(\vx', \eta')|\Omega\rangle
   = \int {d^3k \ov (2\pi)^{3}} Y_{\vk}(\vx)Y_{\vk}^*(\vx)u_k(\eta)u_k^*(\eta') \, ,
 \eeq
 where $|\Omega\rangle $ is the vacuum state.
For convenience, we adopt planar coordinates, i.e. 
$Y_{\vk}(\vx) = e^{i \vk \cdot \vx}$, and, as a shorthand
notation, we call Whightman function also 
 \beq \label{GenWhightman}
  W_k(\eta, \eta') = u_k(\eta)u_k^*(\eta') \, .
 \eeq
The power spectrum, $P_{k}$, is related to the Whightman function. In
particular, at late times
 \beq \label{spectrumtree} 
  P_{k} = {k^3 \ov 2\pi^2} \lim_{\eta \to 0} W_k(\eta, \eta) \, . 
 \eeq 
The mode functions which are entering the expansion of the field are such that $u_k(\eta) = {\mu_k / z}$,
where $\mu_k$ is the solution of the field equation:
 \beq \label{modeequation}
  \mu_k'' +\left(k^2- {z'' \ov z}\right)\mu_k=0.
 \eeq
The quantity, ${z'' / z}$, depends on the background, it is often
called the gravitational source term for   
the seed perturbations~\cite{Mukhanov:1990me}.

In a (quasi) de Sitter case, at leading order in the slow-roll parameters, the
general solution of the mode equation is given by: 
 \beq \label{modesolution}
  u_k(\eta) = {\mu_k \ov z}=
   c_1{\sqrt{-\pi \eta} \ov 2z}H^{(1)}_\nu(-k\eta)+c_2{\sqrt{-\pi \eta} \ov 2z}H^{(2)}_\nu(-k\eta) \, .
 \eeq
where $H^{(1,~2)}$ are the Hankel functions. The flat space solution
(Bunch-Davies vacuum) is recovered for $\eta \to -\infty$, if $c_1=
e^{i\left(\nu + {1 \ov 2}\right){\pi \ov 2}}, \, c_2=0$.\\

In the in-in formalism, the expectation value of any operator
$\widehat{\mathcal{O}}$ at a given time $\eta_0$ is given by:
 \beq \label{ininamplitudes}
  \langle\Omega| \widehat{\mathcal{O}}(\eta_0) |\Omega\rangle =
  \langle\Omega|\bar{T}\left(e^{i\int_{-\infty}^{\eta_0}d\eta H_I}\right) \widehat{\mathcal{O}}(\eta_0) T\left(e^{-i\int_{-\infty}^{\eta_0}d\eta H_I}\right)|
  \Omega\rangle\,.
 \eeq
In the case of the two-point function, expanding in the interaction Hamiltonian leads to: 
 \beq \label{twopointwithcorrections}
  \mathcal{A}(k_1, k_2, \eta) = 
   \mathcal{A}_{\text{tree}}(k_1, k_2, \eta)+
   \mathcal{A}_{\text{loops}}(k_1, k_2, \eta)\,,
 \eeq
 where the correlators, ${\cal A}$, depend on $k_i^2$ and indices are
contracted using the background metric. The time $\eta$ is also 
defined by the unperturbed metric.

\section{Infrared divergences}
  \label{GeneralBehaviour}  

\begin{figure}
\begin{center}
\includegraphics[width=4cm]{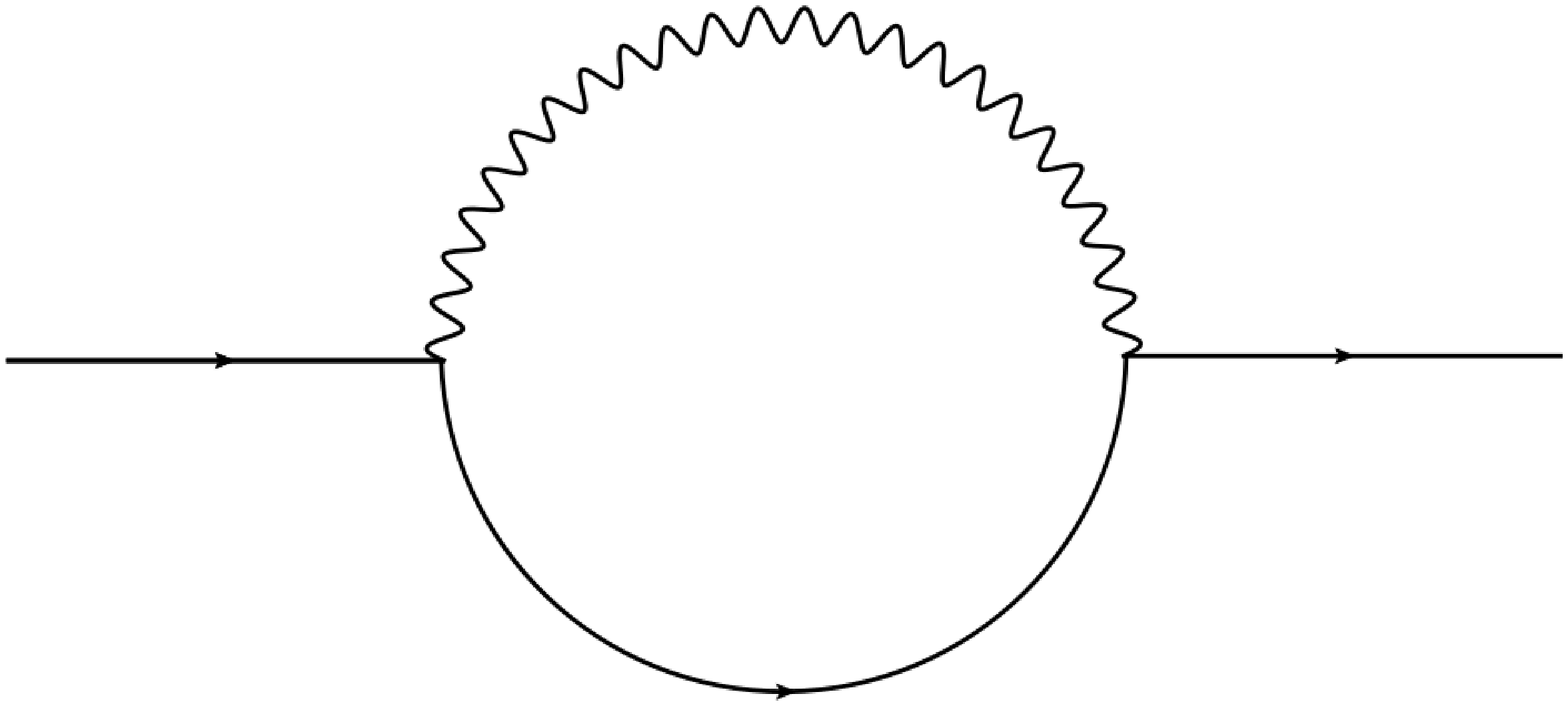}
\end{center}
\caption{Correction to the scalar two point function from an
  intermediate scalar and graviton.} 
 \label{fig1}
\begin{center}
\includegraphics[width=4cm]{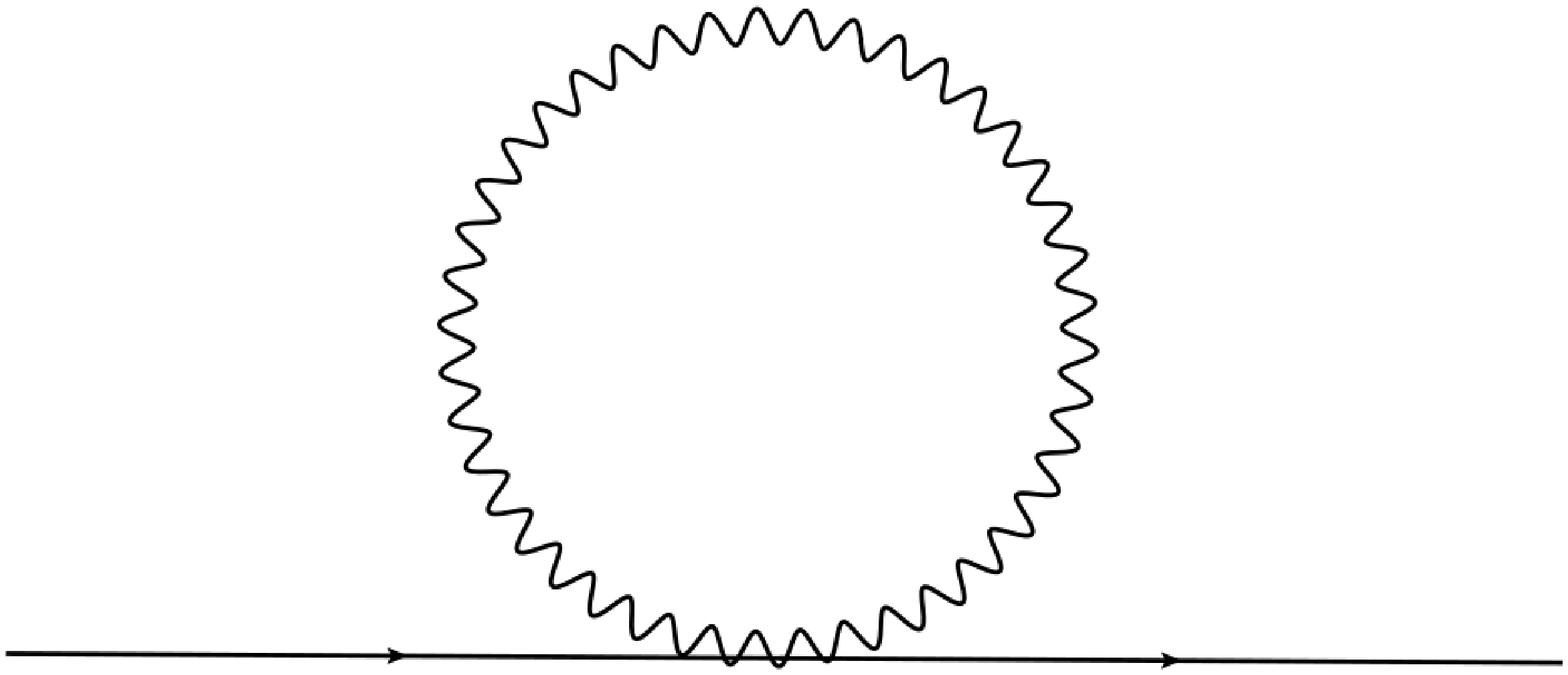}
\end{center}
\caption{Correction to the scalar two point function from a graviton
  bubble.}
 \label{fig2}
\end{figure}

We discuss here the general features of IR divergences using the example of
the lowest-order gravitational loop diagrams that we are typically
interested in: the
self energy diagrams, such as those given
by the cubic interaction vertices, and the ``bubble'' diagram. We present 
these two diagrams in Figs.~(\ref{fig1}) and (\ref{fig2}). The fields running in
the loop can be different, however a generic theory with gravitation
will involve a graviton loop. 

The computation of these diagrams has always been
performed in a pure de Sitter space~\cite{IRLiterature,
  Weinberg:2005vy, Seery:2010kh, Riotto:2008mv,
  Dimastrogiovanni:2008af, Burgess:2009bs, 
Giddings:2010nc,Burgess:2010dd, Koivisto:2010pj}. However, some authors
have also discussed the case of a quasi de Sitter background
\cite{Giddings:2010nc,Koivisto:2010pj}. In order to discuss the IR
divergences, we will not need to know the 
whole expression for these diagrams. Those can be found, for example
in Refs.~\cite{Riotto:2008mv,Dimastrogiovanni:2008af,Burgess:2009bs,Giddings:2010nc}.  
Furthermore, our observations are largely independent of de Sitter or
quasi de Sitter. Let us briefly list here the main 
differences.  In a pure de Sitter case the scalar metric
perturbations are pure gauge, and the only physical perturbations are
the tensor ones, i.e. that of the gravitons. In
Eqs.~(\ref{modeequation},~\ref{modesolution}), $\nu$ becomes equal to
${3 / 2}$ in a pure de Sitter case. 

In a quasi de Sitter case the de Sitter {\em scale invariance} is
broken, which results in a modification of the term ${z'' / z}$ in  
Eq.~(\ref{modeequation}). It is evident from the  experiments that the
breaking is very small, within the observable ranges of scales relevant
for the CMBR measurements, which corresponds to roughly $7$ e-foldings
of inflation.  
It can be parametrized with the slow-roll parameter:
   $ 
    \epsilon = -{\dot H / H^2} \, .
   $ 
The solution to the field  Eqs.~(\ref{modeequation},~\ref{modesolution}) now has
 $ 
  \nu = {3 / 2}+{(1-n)/  2}, 
 $ 
where $n$ is the spectral index~\footnote{Which is a function of
  $\epsilon$ and other small scale breaking parameters. The precise
  values of the spectral   index for gravitons and scalars are 
different, see ~\cite{Mazumdar:2010sa}.}. For a red-tilded spectrum,
as the one observed for scalar perturbations, i.e. $n-1<0$, the IR
divergences of the  
loop integrals become worse than that of the pure de Sitter case. 

The computation of the diagrams is quite involved. Their structure is
not immediately transparent for what concerns the physical
interpretation of the IR divergences. In particular, the
diagrams have two types of integrations - one over the 
(conformal) times for each of the vertices, and the other 
over the loop momentum. There are two possible sources of 
IR ``divergences''. One comes from the time integrals and it is
present only for some kind of interactions \cite{Seery:2010kh}. In our
scenario it will not appear, while, in
theories where it does, it is believed to be cured in realistic models
by using the dynamical
renormalization group techniques~\cite{Burgess:2009bs} (see also 
\cite{Seery:2010kh}). The other IR divergence comes from the
momentum integral and it is always present. We will concentrate on
this latter one.  
 
We briefly discuss the issues regarding the regularization (``box
approach'') of the IR 
divergent integrals. This analysis is actually important to understand
that when the dependence on the cutoffs is not eliminated, the IR
issue is not fully understood, although
the corrections can be made 
sufficiently small. The important point here is that there 
is no unique choice for the IR cutoffs, rendering the results for
the observable quantities ambiguous. 

For instance, the main difference for the IR divergence in the
momentum integration 
arises from the choice of imposing a cutoff on the 
{\em  physical} or on the {\em comoving} momentum. To understand 
its consequences, we consider the
example of a pure de Sitter case and the simplest IR divergent
integral, that for example comes from Fig. \ref{fig2}:
 \beq \label{divergentpart}
  \Lambda^{(\phi)}(\eta) = {1 \ov (2\pi)^3}\int d^3k W_k(\eta, \eta)\,.
 \eeq
By choosing the cutoffs on the {\em physical}
momentum, $p_{UV/IR}  = M_{UV/IR}$, where $M_{UV/IR}$ does
not depend on time, as in \cite{Burgess:2009bs}, one finds that for small $\eta$
 \beq \label{deSitterDivergencePhysicalCutoff}
  \Lambda(\eta) \sim {1 \ov 2\pi^2}\int^{a M_{UV}}_{a M_{IR}} {dk \ov k}
  H^2 = {H^2 \ov 2\pi^2} \log\left({M_{UV} \ov M_{IR}}\right) \, .
 \eeq
The rationale behind this choice is that
this kind of cutoffs does not break de Sitter
scale invariance. 

Instead, choosing the cutoffs on the
comoving wavenumber as, $k_{UV/IR}  = K_{UV/IR}$, where $K_{IR}$ is
independent of time, as in Ref.~ \cite{Giddings:2010nc}, and $K_{UV} =
a(\eta)\mu$, where $\mu$ is 
the renormalization scale, one finds
 \be \label{deSitterDivergenceScaleBreakingCutoff}  
  \Lambda(\eta)  \sim {1 \ov 2\pi^2}\int^{K_{UV}}_{K_{IR}} {dk \ov k}
  H^2 = {H^2 \ov 2\pi^2} \log\left({K_{UV} \ov K_{IR}}\right) \approx {H^2 \ov 2\pi^2}  N_\eta ,
 \ee
where $N_\eta$ is the number of efoldings from the beginning of
inflation up to time $\eta$ and we have neglected the UV contribution
proportional to $\log{\mu \ov H}$. This kind of cutoffs does break de Sitter scale
invariance. The rationale behind this choice is that we integrate over
modes that 
were sub-Hubble at the beginning of inflation, $K_{IR}=a_iH_i$.

We see that Eqs.~(\ref{deSitterDivergencePhysicalCutoff},
\ref{deSitterDivergenceScaleBreakingCutoff}) give very different 
results. First principles do not instruct us to prefer a cutoff
either on the
physical or the comoving momenta\footnote{There are different
  valid physical motivations suggesting different kind of
  cutoffs, for example finiteness of inflation for the comoving one, or causality and/or
  superhorizon scale invariance for the physical one, but there is not
  an utterly univocal 
  principle that truly stands above the others.}, and thus the scale
dependence of the ``observable'' 
changes depending on the cutoff, which clearly makes it
unphysical. The fact that the correlators grow with the
  IR cutoff, that is the ``size of the box'' $L = K_{IR}^{-1}$ also creates
  difficulty in approximating the 
ensemble averages via the spatial averages, which are performed in
practical observations. In fact, the RMS deviation between the spatial
and the ensemble averages goes like 
$\Delta P_k \sim {P_k / (kL)^{3/2}}$ \cite{WeinbergNew}, in the case
of the spectrum, and, since
$P_k$ grows with $L$  due to the IR
divergences of the quantum loops, it could become non-negligible and
should be taken into account for precise measurements and predictions, 
for not too
large boxes and for certain scales (in particular for scales not too
different from the ``box size''). The solution to this problem is to
apply the ergodic theorem only to the true IR-safe correlators.

To resolve the issue of IR divergences,  we must then
propose a recipe for the definition of a sensible perturbation theory
that does not have any of the ambiguities we have presented.

\section{Eliminating IR divergences}
  \label{physicalproposal} 

Our solution of the infrared issue will be centered on the concept of
{\em local observer}, as we said in the introduction. In particular,
our implementation of the principle of locality, will be
different from the one dealt with, for example, in
\cite{Urakawa:2009my, Urakawa:2010kr}, which consists in using gauge
transformations that are singular at large scales. It will also
differ, as we will see, from the approach in 
\cite{Gerstenlauer:2011ti, Byrnes:2010yc}, which, in particular, uses
an explicit cutoff $q_0$, of the size of the typical scale of observations,
when proposing how to resum the contributions from scales larger than
$q_0^{-1}$~\footnote{As we 
  will discuss at the end of section  \ref{amendingprocedure}, this
  proposal is therefore an improved version of the ``physical box
  size'' cutoff technique that we discussed in the introduction:
  effectively the result is equivalent to just using from the start an
  infrared cutoff equal to $q_0$, with
  all the conceptual problems we outlined.}.

Our presentation is divided in three parts: in section
\ref{conceptualpoint} we investigate the physical point at the origin of
the infrared divergences, by discussing what the true observables are
and why
the in-in formalism, as it stands now, does not properly account for
them. In section \ref{IRallorders} we demonstrate that {\em all}
infrared divergences from momentum integrals in gravitational loops 
have a common origin, which is precisely the unphysical
element of the formalism presented in
section \ref{conceptualpoint}. 
Finally, in section \ref{amendingprocedure} we outline our solution of
the infrared issue proposing a more physical redefinition of the in-in
formalism. A detailed example of how our proposal is implemented is given in
section \ref{oneloopexample}.

Our notation will be as follows:
we will write the background FRW metric as $g_{\mu\nu}$:
 \beq \label{backgroundmetric}
  ds^2= g_{\mu\nu}dx^\mu dx^\nu = -dt^2+a^2\delta_{ij}dx^i dx^j.
 \eeq 
and, in our conventions, the background quantities, which are
contracted using the background metric, have no labels: for example
$k^2 = \delta^{ij}k_i k_j$. 

The perturbed quantities are instead indicated by ''over-bar''.
Thus, the local classical metric is given by (neglecting the vector
  part):
 \beq \label{truelocalmetric}
  \overline{ds}^2= 
   \overline g_{\mu\nu}dx^\mu dx^\nu =
    -\overline N^2dt^2+a(t)^2 \overline h_{ij}(dx^i + \overline N^i dt)(dx^j + \overline N^j dt)
 \eeq 
where $N^i, N$ are the shift and lapse functions, and~\footnote{The
  shift and lapse functions are
  Lagrangian multipliers, determined by the constraints as functions of
 the perturbations $\zeta, \gamma$ defined in
 Eq.~(\ref{threemetricperturb}). In Eqs.~(\ref{truelocalmetric}),
 (\ref{threemetricperturb}), we made a  
gauge choice, for what concerns the parametrization of the metric and
the choice of coordinates $(\vec x, t)$. In particular,  
we distinguish scalar and true tensor part, so that
$\gamma_i^i=\partial^i \gamma_{ij} = 0$.  
The choice of a gauge does not introduce any element of
non-physicality or arbitrariness in the 
description of the observable.} 
 \beq \label{threemetricperturb} 
  \overline h_{ij} dx^idx^j= 
   e^{2\zeta(\vec x, t)}(e^{\gamma(\vec x, t)})_{ij} dx^idx^j
  \, . 
 \eeq 
The quantities contracted with this metric will have an ''over-bar'',
for example $\overline k^2 = \overline h^{ij}k_i k_j$.
 
\subsection{True observables and in-in
  formalism} \label{conceptualpoint} 

We will here argue that the usual formalism for the 
perturbation theory does not define correctly the observables from the
point of view of a local observer, and this is the origin of the IR
divergences.

Let us look closely at the definition of observables in the in-in
formalism. For definiteness, let us consider the two-point function as a working example
(our considerations apply also to $N$-point functions, as we will
discuss in section \ref{Npointfunctions}). The definition of the
observables in the perturbative in-in formalism is given by
the right-hand side of Eq.~(\ref{twopointwithcorrections}). The important
element in this formula is that the observables are defined in terms
of the background quantities,  in particular the  
background metric (entering for example in $k^2$). However, 
the background metric has no physical meaning and we claim that
this is the reason of an appearance of the IR divergences.

In fact, any local observer uses true local clocks and ruler. The observer has
no notion of a ``background metric'' and ``perturbations'' on it, but
instead he/she uses the local metric for his/her measurements. We will
show that when writing the perturbative correlators from the point of view
of the local observer, that is using the local metric, the IR
divergences {\it are} cured.

\begin{figure}
\begin{center}
\includegraphics[width=8cm]{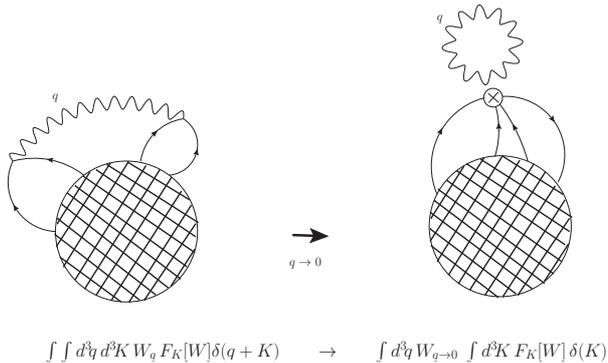}
\end{center}
\caption{A schematic representation of the loop
  corrections in an IR limit as shown in the text. A generic correlator can be written,
  taking care of the time-ordering, as a
  product of Whigtman functions integrated over the momenta running in
  the internal lines, with delta functions at vertices. In the IR
limit of a diagram line (for example, here in the picture, a
graviton's one), the total correlator factorizes
into the product of a disconnected two-point function at small
momentum times the rest of the correlator with a certain
``vertex''. The disconnected two-point function
(when $q\rightarrow 0$) leads to the IR divergences.
We will show that when redefining the correlators using the local
metric, the disconnected contributions disappear.} 
 \label{disconnectedbubble}
\end{figure}

In more details, the IR divergences arise for the following reason. In
the in-in formalism, the perturbations ($\zeta, \gamma$) are promoted 
to quantum fields ($\widehat\zeta, \widehat\gamma$), and
path-integrated using the background metric of Eq.~(\ref{backgroundmetric})
to contract the indices. We will show that in the correlators at loop
level, {\em all} IR divergences arise when at least one of the momenta
running in the loops lines goes to zero. In that limit the correlator becomes effectively {\it
  disconnected} and leads to the IR divergences.  
We represent this, schematically, in Fig.~\ref{disconnectedbubble}.  

We claim that it is precisely these operations of expansion over the
background metric and quantum averaging of the perturbations in effectively
disconnected contributions that make no sense from the 
point of view of an observer.
These are the unphysical elements of the standard in-in perturbation
theory, because a local observer will not be sensitive to the
background metric Eq.~(\ref{backgroundmetric}), but to the local
metric Eq.~(\ref{truelocalmetric}), whose 
classical value in the quantum theory is
 \beq \label{classicallocalmetric}
  \overline g_{\mu\nu} = \langle\, \widehat{\overline g}_{\mu\nu} \,\rangle \, ,
 \eeq 
where, in particular,
 \beq \label{classicallocalthreemetric}
  \overline h_{ij} = 
   \langle e^{2\widehat\zeta(\vec x, t)}(e^{\widehat\gamma(\vec x, t)})_{ij} \rangle
  \, . 
 \eeq

We will show in section \ref{amendingprocedure} that by correctly
defining the correlators in terms of 
the local classical metric Eqs.~(\ref{truelocalmetric}),
(\ref{classicallocalthreemetric}) these {\it disconnected} pieces
will be removed and the infrared issue cured. Our IR-safe correlators
will be different from those defined in other proposals, as for
example in \cite{Gerstenlauer:2011ti, Byrnes:2010yc} or
\cite{Urakawa:2009my, Urakawa:2010kr}. 


\subsection{IR divergences at all order}\label{IRallorders}

We will now prove that indeed {\em all} IR divergences from momenta
integrals originate in the
unphysical expansion of the local metric
in perturbations and their path 
integration in the IR limit
where higher order correlators become disconnected, as we have claimed
in the previous section.
Here, we will focus on the two-point function for scalar perturbations, but our analysis 
can be extended to the general case of $N$-point functions and to
gravitons. 

Let us first review in some detail the path-integral formulation of the in-in
perturbation theory, for a general reference see \cite{Weinberg:2005vy}. The
Lagrangian we are keen to discuss is the one of a  
nearly massless scalar field, $\sigma$, which could be the inflaton,
minimally coupled to gravity, 
 \beq\label{actionsigma}
  S= {1 \ov 2}\int\sqrt{-\text{det}(\overline g)}
  \left[\overline R-\overline g_{\mu\nu}\partial_{\mu}\sigma\partial_{\nu}\sigma-2V(\sigma)\right] \, ,
 \eeq
where $V(\sigma)$ is very flat, otherwise arbitrary. Using the ADM
formalism~\cite{Arnowitt:1962hi}, the action becomes
\begin{multline}
 S = {1 \ov 2}\int\sqrt{-\text{det}(\overline h)}~
 {a^3 \ov 2}~e^{3\zeta}\Bigg[\overline N\overline
   R^{(3)}-2\overline NV(\sigma)+\overline N^{-1}\Big(\overline E^j{}_i\overline E^i{}_j-(\overline E^i{}_i)^2\Big)+
    \nonumber\\
  + \overline N^{-1}\Big(\dot\sigma-\overline h^{ij} \overline N_i\partial_j\sigma\Big)^2-
    \overline Na^{-2}e^{-2\zeta}[\exp{(-\gamma)}]^{ij}\partial_i\sigma\partial_j\sigma\Bigg]
  \;,
\end{multline}
where 
 \beq \label{tensorEij}
   \overline E_{ij}\equiv \frac{1}{2}\Big(\dot{\overline h}_{ij}-\nabla_i \overline N_j-\nabla_j\overline N_i\Big)\;;
 \eeq
$\nabla_i$ is the
three-dimensional covariant derivative calculated with the
three-metric $\overline h_{ij}$; and $\overline R^{(3)}$ is the curvature scalar calculated with
this three-metric: 
$$
\overline R^{(3)}=a^{-2}e^{-2\zeta}\Big[e^{-\gamma}\Big]^{ij}~\overline R^{(3)}_{ij}.
$$
All spatial indices $i$, $j$, etc. are lowered  and raised  with the
metric $\overline h_{ij}$ and its reciprocal. In particular,
by computing the derivatives, such as 
$\dot{\overline h}_{ij}$ in Eq.~(\ref{tensorEij}), one can find a kinetic
term for $\zeta$ and $\gamma$, where indices are contracted by
$\overline h_{ij}$. 
We choose the gauge
where the scalar field is homogeneous, and $\zeta \neq 0$.

The two-point function in the standard in-in
perturbation theory (using the
path-integral formulation) is obtained by expanding the metric
$\overline{h}_{ij}$ in terms of the perturbations $\zeta, \gamma$. In
this section we will not indicate quantum fields with 
$\;\widehat{\mbox{}}\;$ in order not to clutter the equations. The
formalism also requires doubling the field degrees of freedom in order
to account for the time ordering and anti-ordering in
Eq.~(\ref{ininamplitudes}): 
$\zeta \to \{\zeta_+, \zeta_-\}$, and similarly for $\gamma$. It will
appear to be
convenient to change basis to
$\zeta_C = {1 \ov 2}(\zeta_+ + \zeta_-)$,
$\zeta_\Delta = \zeta_+ - \zeta_-$, and the analogous for the $\gamma$.

The
perturbative expansion in terms of the perturbations $\zeta, \gamma$
is obtained by functionally Taylor-expand the exponential 
$e^{i\int_{\eta_{\text{in}}}^{\eta}d\tau \int d^3x \sqrt{-det\,\overline h} \, \mathcal{L}(\zeta_C,\gamma_C,\zeta_\Delta,\gamma_\Delta)}$ 
in the path integral in powers of the perturbations
$\zeta_C,\gamma_C,\zeta_\Delta,\gamma_\Delta$: 
 \be \label{twopointininlagrangian}
  \int {d^3 k_1 \ov (2\pi)^3} & \int {d^3 k_1 \ov (2\pi)^3} 
   e^{i\vec k_1\cdot \vec x_1 + i\vec k_2\cdot \vec x_2}\langle\zeta(x_1, \eta) \zeta(x_2, \eta)\rangle = \nonumber \\
   & \int {d^3 k_1 \ov (2\pi)^3}\int {d^3 k_1 \ov (2\pi)^3} 
   e^{i\vec k_1\cdot \vec x_1 + i\vec k_2\cdot \vec x_2}
   \int_{\mathcal{C}} \mathcal{D}\zeta_C\mathcal{D}\zeta_\Delta\mathcal{D}\gamma_C\mathcal{D}\gamma_\Delta
    \, \zeta_C(x_1, \eta) \zeta_C(x_2, \eta) \, \nonumber \\
   & \sum_{\substack{n, m, \\ u, v \geq 0}} {1 \ov n! m! u! v!} 
   \prod_{a=0}^n\biggl(\int d^3x_{(a)}d\eta_{(a)}\zeta_C\text{{\small$(\vec x_{(a)}, \eta_{(a)})$}}{\delta \ov \delta  \zeta_C\text{{\small$(\vec x_{(a)}, \eta_{(a)})$}}}\biggr) \nonumber \\
   & \qquad\prod_{b=0}^m\biggl(\int d^3x_{(b)}d\eta_{(b)} \gamma_{\text{{\tiny$Ci_{(b)}j_{(b)}$}}}\text{{\small$(\vec x_{(b)}, \eta_{(b)})$}}{\delta \ov \delta \gamma_{\text{{\tiny$Ci_{(b)}j_{(b)}$}}}\text{{\small$(\vec x_{(b)}, \eta_{(b)})$}}}\biggr) \nonumber \\
   & \qquad\prod_{c=0}^u\biggl(\int d^3x_{(c)}d\eta_{(c)} \zeta_\Delta\text{{\small$(\vec x_{(c)}, \eta_{(c)})$}}{\delta \ov \delta \zeta_\Delta\text{{\small$(\vec x_{(c)}, \eta_{(c)})$}}}\biggr) \nonumber \\
   & \qquad\prod_{d=0}^v\biggl(\int d^3x_{(d)}d\eta_{(d)} \gamma_{\text{{\tiny$\Delta\,i_{(d)}j_{(d)}$}}}\text{{\small$(\vec x_{(d)}, \eta_{(d)})$}}{\delta \ov \delta \gamma_{\text{{\tiny$\Delta\,i_{(d)}j_{(d)}$}}}\text{{\small$(\vec x_{(d)}, \eta_{(d)})$}}}\biggr) \nonumber \\
   & \quad\qquad  \;\;
   e^{i\int_{\eta_{\text{in}}}^{\eta}d\tau \int d^3x
     \sqrt{-det\,\overline h} \,
     \mathcal{L}(\zeta_C,\gamma_C,\zeta_\Delta,\gamma_\Delta, \overline h_C, \overline h_\Delta)}
   \Psi_{\text{vac}, C}(\eta_{\text{in}})\Psi_{\text{vac}, \Delta}(\eta_{\text{in}})
  \, , 
 \ee
where $\mathcal{C}$ is the closed time path, defined by
$\zeta_+(\eta_{\text{in}})= \zeta_-(\eta_{\text{in}})$ (similarly for other fields), and the vacuum
function is $\Psi_{\text{vac}, C/\Delta}$. Finally, 
 $ 
  \mathcal{L}(\zeta_C,\gamma_C,\zeta_\Delta,\gamma_\Delta, \overline h_C, \overline h_\Delta) =
  \mathcal{L}(\zeta_+,\gamma_+, \overline h_+)-\mathcal{L}(\zeta_-,\gamma_-, \overline h_-),
 $ and, {\em after} the functional derivations,
  $\zeta_C,\gamma_C,\zeta_\Delta,\gamma_\Delta \to 0$,
  $\overline h_{ij} \to \delta_{ij}$, in 
  $\mathcal{L}$ for the usual Taylor expansion around the
  unperturbed background. 

We are now ready to investigate what kind of IR divergences from
momentum integrals are present in the
cosmological perturbation theory in a (quasi) de Sitter background. 

We find that {\em all} these IR divergences, at all
orders, appear when a momenta running in a loop line goes to zero, as we
claimed. The proof of this is as follows: in the perturbation theory over (quasi) de
Sitter, the Whightman functions go as~\footnote{More precisely, in
  the (quasi) de Sitter case, they would go as 
$\sim |\vec p|^{-(4-n)}$, where $n$ is the spectral index.}:
 \beq
  \sim {1 \ov |\vec p|^3} = {1 \ov |\vec k-\vec q|^3} 
   = {1 \ov (|\vec k|^2+|\vec q|^2-2|\vec k||\vec q|\cos{\theta})^{3 \ov 2}}\,,
 \eeq
and therefore this would diverge only when $\vec p = 0$. This happens when
the momentum running in the line goes to zero.

In particular, there are no {\it collinear} divergences arising when two
momenta are parallel, because the denominator of the Whightman
functions (from which all other propagators can be obtained) never
goes to zero in that case. This is very different from what happens
in a scattering amplitude on Minkowski background, where such
collinear divergences are present, for example when a soft gluon (or
graviton) line originates from a massless line yielding a
contribution to the diagram of the form:
$\sim {1 \ov p^\mu p_\mu} = {1 \ov (k-q)^\mu (k-q)_\mu} = {1 \ov -2k^\mu q_\mu}$, which
diverges when $k^\mu$ is parallel to $q^\mu$.

The IR divergences in momentum loops in perturbation theory 
can therefore all be  
recovered by taking the small momentum limit of the propagators in the
relevant loop lines. Now we want to prove the other claim of ours:
that all the IR divergences appear from 
disconnected contributions to the correlators. The relevant
propagators in the usual $\pm$ basis are: 
\be
 G^{-+}(\eta, \eta') & =  W(\vx, \eta, \vx', \eta') \,,
 \qquad
 G^{+-}(\eta, \eta') =  W(\vx, \eta', \vx', \eta) \,, \nonumber \\
 G^{++}(x,y) & = \theta(\eta-\eta') G^{-+}(\eta, \eta')
 + \theta(\eta'-\eta) G^{+-}(\eta, \eta') \,, \\
 G^{--}(x,y) & =  \theta(\eta-\eta') G^{+-}(\eta, \eta')
 + \theta(\eta'-\eta) G^{-+}(\eta, \eta') \,,\nonumber
\ee
where $W(\vx, \eta, \vx', \eta')$ is the Whightman function (recall
Eq.~(\ref{WhightmanfunctionModeexpansionState})). Note that 
$G^{++} + G^{--} = G^{+-} + G^{-+}$.
In the new, more convenient, ($C, \Delta$) basis the correlation functions are
\beq
 \left(\begin{array}{cc}
  G_C & G_R \\
  G_A & 0 \\
 \end{array}\right)
 = Q \left( \begin{array}{cc}
  G^{++} & G^{+-} \\
  G^{-+} & G^{--} \\
 \end{array} \right) Q^T \,,
 \qquad
 Q = \left( \begin{array}{rrr}
  {1 \ov 2} && {1 \ov 2} \\
  1 && -1 \\
 \end{array} \right) \,.
  \label{Wmatrix}
\eeq
The advanced and retarded propagators are related by $G_A(x,y)
= G_R(y,x)$ and vanish in the coincidence limit. The convenience in
using this basis descends from the fact that in the 
IR limit (small momentum), the propagators behave as follows
 \beq
  G_C(q \equiv |\vec q|, \eta, \eta') \underset{q \to 0}{\simeq}
   W(q, \eta, \eta')|_{q \to 0} \sim { H^2 \ov q^{(4-n)}}\,, \qquad
  G_R(q, \eta, \eta')  \underset{q \to 0}{\sim} \theta(\eta-\eta')
   {(\eta^{2\nu}-\eta^{'\,2\nu}) \ov (\eta \eta)^{'\,{{3 \ov 2}-\nu}}}
 \eeq
and therefore the IR divergences coming from the vanishing of
the momentum in loop lines are accounted for by the $G_C$ propagator
only. We concentrate therefore on the expansion in $\zeta_C, \gamma_C$.

Let us then see how all these IR divergences arise from disconnected
contributions to the correlators. It is easy to do it now
that we have shown that they all arise when one (or more)
momenta in a diagram line
goes to zero. Indeed, the small momentum limit of the propagators can be
found using the expansion of the perturbation fields
given by Eq.~(\ref{fieldexpansion}), where - we recall - we have chosen
planar coordinates: $Y_{\vk}(\vx) = e^{i \vk \cdot \vx}$.
In this limit, $Y_{\vk}(\vx) \sim 1$, and thus the momenta of
the lines sent to this IR limit drop out of the
delta functions at the interaction vertices in
Eq.~(\ref{twopointininlagrangian}). Therefore, in the IR limit
the contributions of these lines to the two-point
function become disconnected, and
Eq.~(\ref{twopointininlagrangian}) reads, once written in momentum space, as:
 \begin{multline} \label{twopointininlagrangiandecoupled}
  \!\!\!\!\langle \zeta(k_1, \eta) \zeta(k_2, \eta)\rangle_{\text{IR, all orders}} \!\!= \!\!
   \sum_{n, m \neq (0, 0)} {1 \ov n! m!} \!\!
  \prod_{\substack{0 \leq a \leq n \\ 0 \leq b \leq m}}\biggl(\!\int d\eta_{(a)} d\eta_{(b)}\!\biggr)
  \langle \prod_{a=0}^n \zeta_C(\eta_{(a)})\rangle_{_\text{(IR)}} 
  \langle\prod_{b=0}^m \gamma_{C\,i_{(b)}j_{(b)}}(\eta_{(b)})\rangle_{_\text{(IR)}} \\
  \int_{\mathcal{C}}\mathcal{D}\zeta_C\mathcal{D}\zeta_\Delta\mathcal{D}\gamma_C\mathcal{D}\gamma_\Delta
    \, \zeta_C(k_1, \eta) \zeta_C(k_2, \eta) \,  
   \prod_{a=0}^n\biggl(\int d^3x_{(a)}{\delta \ov \delta \zeta_C(\vec x_{(a)}, \eta_{(a)})}\biggr) \\
   \prod_{b=0}^m\biggl(\int d^3x_{(b)}{\delta \ov \delta \gamma_{Ci_{(b)}j_{(b)}}(\vec x_{(b)}, \eta_{(b)})}\biggr)
  \, 
   \, e^{i\int_{\eta_{\text{in}}}^{\eta}d\tau \int d^3x \sqrt{-\text{det}\overline h} \, \mathcal{L}(\zeta_C,\gamma_C,\zeta_\Delta,\gamma_\Delta, \overline h_C, \overline h_\Delta)}
  \Psi_{\text{vac}, C}(\eta_{\text{in}})\Psi_{\text{vac}, \Delta}(\eta_{\text{in}}) \, , 
 \end{multline}
where again, having Taylor expanded,
  $\zeta_C,\gamma_C,\zeta_\Delta,\gamma_\Delta \to 0$,  $\overline h_{ij} \to \delta_{ij}$, in 
  $\mathcal{L}$ after the functional derivations. The label
$_{_\text{(IR)}}$ indicates the small momentum limit of the
propagators (infrared divergent).

We have thus proven that the equation (\ref{twopointininlagrangiandecoupled}) 
comprises all the IR-divergent parts of the two-point correlator from
loop momenta integrations, and
these appear as disconnected contributions.
With a more schematic and shortcut notation, we can write
Eq.~(\ref{twopointininlagrangiandecoupled}) as 
 \begin{multline} \label{compacttwopointIR}
  \langle \zeta(k_1, \eta) \zeta(k_2, \eta)\rangle_{\text{IR, all orders}} = \\
   \sum_{n, m \neq (0, 0)} {1 \ov n! m!} \,
   \langle \prod_{u=0}^n\zeta\rangle_{\text{{\tiny IR}}} \,
   \langle\prod_{v=0}^m\gamma_{i_{(v)}j_{(v)}}\rangle_{\text{{\tiny IR}}} \, 
    {\partial^n \ov \partial \zeta_{_{\bar h}}^n} \,
    {\partial^m \ov \partial \gamma_{_{\bar h} i_{(v)}j_{(v)}}^m}
   \,\, \langle \zeta(k_1, \eta) \zeta(k_2, \eta)\rangle \, .
 \end{multline}
With
the suffix $\mbox{}_{_{\bar h}}$ we indicate that the derivatives
do not act on the external fields, $\zeta(k_1, \eta), \zeta(k_2, \eta)$,
of the correlator, but on the
fields entering the correlator via its dependence on 
$\overline{h}_{ij} = e^{2 \zeta}(e^{\widehat \gamma})_{ij}$. 

The form of Eq.~(\ref{compacttwopointIR}) makes it even more
evident that the IR divergences are a  
result of the expansion of the local perturbed metric over the
long-wavelength perturbations and the path integration in 
the limit where higher order correlators become disconnected. 
This result completes and extends that in Ref.~
\cite{Giddings:2010nc}, where it was shown 
that some IR divergences in certain one-loop in-in computations could be
recovered from semiclassical expansions around the perturbed
metric~\footnote{In Ref.~\cite{Byrnes:2010yc}, it was also
  argued that some loop IR-divergence could be obtained in this way,
  but the $\delta N$ formalism was used instead, which was questioned
  in \cite{Giddings:2010nc}.}.
We have here proven, using a rigorous field theory
formalism, that all the IR divergences in momentum integrals in
gravitational loops at all orders are accounted for by the 
equation (\ref{compacttwopointIR}). 
The demonstration applies to $N$-point function for all $N$'s as well. 

Since now we see that all IR
divergences in gravitational loops have the same origin, whose nature
we can understand, we will propose a procedure to fully resolve them on
the basis of the physical concepts discussed in section
\ref{conceptualpoint}. In the next section we turn to this point. 


\subsection{Defining a different in-in perturbation theory}
\label{amendingprocedure}

We set out now to define a new perturbative expansion in the in-in
formalism, capable of curing the infrared issue. We will outline our recipe, and
comment on how it differs from other proposals concerning the IR
issue. For definiteness, we will be illustrating 
our proposal on the two-point correlator $  \langle \zeta(k_1, \eta)
\zeta(k_2, \eta)\rangle$, but, just as before, our considerations also apply 
to $\langle \gamma \gamma\rangle$, or any other correlator of $N$
fields, for all $N$'s.

In the usual perturbation theory around the background metric, the
two-point correlator is given by a series
 \beq \label{notationtwopoint} 
  \langle \zeta(k_1, \eta) \zeta(k_2, \eta)\rangle =
  \mathcal{A}(k, \eta)\delta(\vec k_1+\vec k_2) =  
  \bigl(\mathcal{A}(k, \eta)_{_\text{tree}} +
   \mathcal{A}(k, \eta)_{_\text{loops}}\bigr) \delta(\vec k_1+\vec k_2)
 \eeq
We stress that we do not take any infrared limit of sorts here, but
instead we are considering the full perturbative series
($\mathcal{A}(k, \eta)_{_\text{loops}} =\mathcal{A}(k,
\eta)_{_\text{1-loop}}+\mathcal{A}(k, \eta)_{_\text{2-loop}}+
\ldots$) with the loops contributions including the ultraviolet, finite and
infrared parts altogether (as we are not concerned here with the
ultraviolet divergences, we will assume that all correlators here
and in the following are suitably renormalized). 

The background metric enters in this formula because it is used to
contract the comoving momenta as (recall that we use the ADM formalism)
 \beq
   k^2 = h_{_\text{bckgr}}^{ij} \,k_j k_j = \delta^{ij} \,k_j k_j
 \eeq
where we have used $h_{_\text{bckgr}}^{ij} = \delta^{ij}$.

Motivated by our results and understanding in the previous sections,
we set out to define a new perturbation theory, where scales are not
defined by contracting with the
background metric, but with the field
 \beq \label{classicaleffectivethreemetric}
  \overline h_{ij} = 
   \langle e^{2\widehat\zeta(\vec x, t)}(e^{\widehat\gamma(\vec x, t)})_{ij} \rangle
  \, . 
 \eeq
The idea, let us repeat it once again,
is that in this way we should be able to take into account the
actual local metric that defines scales for the observer. 
We want to see if in
this way the infrared issue will be cured or not.
Please, observe that in this definition, the notation $\langle ~~
\rangle$ indicates a quantum expectation value in the in-vacuum and
that $\widehat\zeta, \,\widehat\gamma$ are the full quantum
fields. Therefore the definition of $\overline h$ is the standard
definition of the classical field as the in-vacuum quantum average of
a quantum operator. In 
particular, it is not based on any infrared limit or large scale
averaging or any infrared/large scale
definition/quantity/procedure.  

As the background metric was entering the correlators by contracting
the wavenumbers $k$, the new field $\overline h$ (the local metric)
will have to be used 
to contract the wavenumbers in place of the background metric, as
   \beq \label{replacement}   
     \overline k^2 =(\overline{h})^{ij} \,k_j k_j
     = \langle e^{-2\widehat \zeta(\vec x, t)}(e^{-\widehat \gamma(\vec x, t)})^{ij}\rangle\; k_i k_j . 
   \eeq
Note that this quantity is very different from the one defined in
Refs~\cite{Gerstenlauer:2011ti, Byrnes:2010yc}, which is
 \beq \label{kappacut}
  \kappa(q_0, L)^2 = e^{-2\zeta_{ir}(\vec x, t)}(e^{-\gamma_{ir}(\vec x, t)})_{ij}k^i k^j \, ;
 \eeq 
the latter quantity is in fact defined by using the large scale behavior
 \beq \label{infraredfieldstasetal}
    \zeta_{ir}/\gamma_{ir} = (2\pi)^{-{3 \ov 2}}\underset{L^{-1}< q \ll q_0}{\int} d^3q   
    e^{i \vec k \cdot \vec x} \zeta_{\vec k}/\gamma_{\vec k} \; ,
 \eeq
where $q_0$ is a suitable cutoff defining an infrared/large scale
limit where the fields become classical\footnote{In
  Refs~\cite{Gerstenlauer:2011ti, Byrnes:2010yc}, $q_0$ is
the scale of observation. There is a similar quantity -defined however
in position space- in \cite{Urakawa:2010kr} where $q_0$ is given
by the Hubble rate.}. As we said, we use instead the full quantum
fields and the usual quantum vacuum expectation value, see equation
(\ref{classicaleffectivethreemetric}). The
logic at the basis of the proposal in
Refs~\cite{Gerstenlauer:2011ti, Byrnes:2010yc} and of equation
(\ref{kappacut}) is the definition of a different background,
characterized by the scale $q_0$, taking into account superhorizon
modes, on which to
do a new perturbation theory. Their power spectrum is defined as
the Fourier transform of a spatial average at two points 
$\vec x_1$ and $\vec x_2 =\vec x_1 + \vec y$ with $q_0^{-1} \sim y$ and
then distances are rescaled with the new background metric, which is
not fully local as well (it depends on $q_0^{-1}\sim |\vec x_2 -\vec x_1|$).

We, instead, put the
physical concept of locality at the basis of our redefinition of the
perturbation theory, asking what a local observer would actually
use: (\ref{classicaleffectivethreemetric}) is not a background
metric, but the local metric that the observer would use to define
scales and a local quantities. Further differences among the
other proposals and ours will soon appear. 

At first sight, at this point we can construct two kind of correlators
using $\overline h$ that we might think to use to define a new
perturbation theory: 
  \beq \label{perturbcorrtrasf}
  \mathcal{A}'(\overline k, \eta) =
   \mathcal{A}'(\overline k, \eta)_{_\text{tree-level}} +
   \mathcal{A}'(\overline k, \eta)_{_\text{loops}} \, ,
 \eeq
or
  \beq \label{perturbcorrnontrasf}
  \mathcal{A}(\overline k, \eta) =
   \mathcal{A}(\overline k, \eta)_{_\text{tree-level}} +
   \mathcal{A}(\overline k, \eta)_{_\text{loops}} \, .
 \eeq
Here, the first object $\mathcal{A}'(\overline k, \eta)$ is the 
transformed two-point correlator under the transformation $k \to
\overline k$ acting as a change of coordinates. The second object
$\mathcal{A}(\overline k, \eta)$ is instead the {\em non-transformed}
correlator evaluated on the newly contracted $\overline k$. We claim
that the latter is the one we should use for the new perturbation
theory. 

In fact, it is straightforward to see that (\ref{perturbcorrtrasf})
does not give any new perturbation theory, because 
$\mathcal{A}'(\overline k, \eta) =   \mathcal{A}(k, \eta)$, since the
correlators transform as scalar under the transformation $k \to
\overline k$, and thus we would be just re-writing in new coordinates the
same old perturbation theory, with all the same problems.

Instead $\mathcal{A}(\overline k, \eta)$ is obviously different from
$\mathcal{A}(k, \eta)$ and thus defines a new perturbation theory.
It is straightforward to obtain the explicit expression of
$\mathcal{A}(\overline k, \eta)$ in terms of $\mathcal{A}(k, \eta)$,
by using standard techniques in calculus and the action
of diffeomorphisms. We briefly illustrate the passages:
\begin{itemize}
 \item we take the
correlator $\mathcal{A}(k, \eta)$ (which comprises the tree-level and
all loop contributions, see equation (\ref{notationtwopoint})) and
transform its $k$-dependence into a 
$k(\overline k)$-dependence using
   \beq \label{replacementinverse}   
     k^2 =(\overline{h}^{-1})^{ij} \,\overline k_j \overline k_j
     = \langle e^{2\widehat \zeta(\vec x, t)}(e^{\widehat \gamma(\vec x, t)})^{ij}\rangle\; \overline k_i \overline k_j , 
   \eeq
 \item we then expand in Taylor series:
 \beq \label{expandamplitudeoverlinek} 
   \mathcal{A}(k(\overline k), \eta) =
   \mathcal{A}(\overline k, \eta)\big|_{\zeta_{_{\bar h}}=\gamma_{_{\bar h}} = 0}
   + \!\!\!\!
   \sum_{\substack{n, m \geq 0 \\ \{n, m\} \neq \{0, 0\}}} {1 \ov n! m!} \langle(\widehat\zeta)^n\rangle \langle(\widehat\gamma^{ij})^m\rangle
   \partial_{\zeta_{_{\bar h}}}^n  \partial_{\gamma^{ij}_{_{\bar h}}}^m 
    \mathcal{A}(k, \eta)\big|_{\substack{\zeta_{_{\bar h}}=\gamma_{_{\bar h}} = 0 \\ (k^2 = \overline k^2)}}\, ,
 \eeq
 \item we finally note that the first term
 on the right hand side of Eq.~(\ref{expandamplitudeoverlinek}), i.e.
 \be \label{IRsafefirstway}  
   \mathcal{A}(\overline k, \eta)\big|_{\zeta_{_{\bar h}}=\gamma_{_{\bar h}} = 0} & =
   \mathcal{A}(k(\overline k), \eta) - \!\!\!\!
   \sum_{\substack{n, m \geq 0 \\ \{n, m\} \neq \{0, 0\}}} {1 \ov n! m!} \langle(\widehat\zeta)^n\rangle \langle(\widehat\gamma^{ij})^m\rangle
   \partial_{\zeta_{_{\bar h}}}^n  \partial_{\gamma^{ij}_{_{\bar h}}}^m 
    \mathcal{A}(k, \eta)\big|_{\substack{\zeta_{_{\bar h}}=\gamma_{_{\bar h}} = 0 \\ (k^2 = \overline k^2)}}\,  
 \ee
is precisely the non-transformed
correlator evaluated on $\overline k$, which we were looking for. Once
again, we stress that we have not taken any infrared limit or similar,
and that all higher order contribution include all ultraviolet, finite
and infrared parts.
\end{itemize}
 
In fact, at this point it is not yet clear that the infrared issue has been
cured in this way. In order to understand that, we need to look more
carefully at the loops contribution in 
$\mathcal{A}(\overline k, \eta)$ to assess better how the new
formalism, centered around $\overline h$, differs from the old one
centered around the background metric.
Thus, by substituting (\ref{notationtwopoint}),
(\ref{perturbcorrtrasf}) in equation (\ref{IRsafefirstway}), we find
that~\footnote{Here we have also used 
$\mathcal{A}'(\overline k, \eta) = \mathcal{A}(k, \eta)$ to write
  everything more neatly in terms of $\overline k$.}
 \beq \label{IRsafefirstwayloops}
  \mathcal{A}(\overline k, \eta)_{_\text{loops}} =
   \mathcal{A}'(\overline k, \eta)_{_\text{loops}} -
   \sum_{\substack{n, m \geq 0 \\ \{n, m\} \neq \{0, 0\}}} {1 \ov n! m!} \langle(\widehat\zeta)^n\rangle \langle(\widehat\gamma^{ij})^m\rangle
   \partial_{\zeta_{_{\bar h}}}^n  \partial_{\gamma^{ij}_{_{\bar h}}}^m 
    \mathcal{A}(k, \eta)\big|_{\substack{\zeta_{_{\bar h}}=\gamma_{_{\bar h}} = 0 \\ (k^2 = \overline k^2)}}\, .
 \eeq
Clearly, $\mathcal{A}(\overline k, \eta)_{_\text{loops}}$ is different
from zero, because the first 
term
on the right hand side
is not equal to the the second term (the sum over disconnected
contributions), and thus their difference is not
null. In fact the first term is the standard loops contribution, not
given by disconnected diagrams. Recall that all terms at each order include
their ultraviolet, finite and infrared parts altogether.

However, because of our results in section \ref{IRallorders}, we see
that {\em the infrared 
  divergent parts} do cancel between the first and the second term on
the right hand side of equation
(\ref{IRsafefirstwayloops})\footnote{Note the crucial 
minus sign in front of all the higher order terms in our equation
(\ref{IRsafefirstway}). Such minus has nothing
  to do with an expansion in $k$ of
 $\overline k^2 = 
\langle e^{-2\widehat \zeta(\vec x, t)}(e^{-\widehat \gamma(\vec x, t)})^{ij}\rangle\; k_i k_j$, 
which would instead give
 \beq \label{expandamplitudek} 
   \mathcal{A}'(\overline k(k), \eta) =
   \mathcal{A}(k, \eta)\big|_{\substack{\zeta_{_{\bar h}}=\gamma_{_{\bar h}} = 0 \\ (k^2 = \overline k^2)}}
   + 
   \sum_{n, m \neq \{0, 0\}} {1 \ov n! m!} \langle(\widehat\zeta)^n\rangle \langle(\widehat\gamma^{ij})^m\rangle
   \partial_{\zeta_{_{\bar h}}}^n  \partial_{\gamma^{ij}_{_{\bar h}}}^m 
    \mathcal{A}(k, \eta)\big|_{\substack{\zeta_{_{\bar h}}=\gamma_{_{\bar h}} = 0 \\ (k^2 = \overline k^2)}}\,.
 \eeq
because Wick's
theorem always forces $n$ and $m$ to be even in order to have non-zero
correlators. This also shows that (\ref{IRsafefirstway}) is not
equivalent to a gauge/coordinate transformation.} and thus 
$\mathcal{A}(\overline k, \eta)$ is free of infrared
divergences: looking at equation (\ref{compacttwopointIR}) it is  
straightforward to see that infrared divergences/Log-enhancements,
{\em and only those}, do cancel out exactly order by order.

We stress that, instead, the ultraviolet (renormalized) and
finite terms do not cancel between the first and second term and so
the loop contributions 
$\mathcal{A}(\overline k, \eta)_{_\text{loops}}$ in the new
perturbation theory around $\overline h$ are 1) different
from those in the standard perturbation theory using the background
metric (thus the new perturbation theory is a non-trivial modification
and in principle testable),
and 2) have the bonus that the IR divergences are absent (without
using any IR cutoff procedure to define the new theory).

As a comment, let us observe that, being fully expressed as a
combination of usual standard coordinate-transformed correlators, as
visible in equation (\ref{IRsafefirstway}),  also the new
correlators will be sharing the same properties of the old ones (such
as conservation and alike) at
large scales (large $\overline k$). Furthemore, since the field $\zeta$
is fully gauge invariant 
and, at the level of field equations,
(\ref{replacement}) acts as a coordinate transformation,
$\zeta$ will be conserved on superhorizon $\overline k$-scales.
Note also that we could have obtained our result using
different gauge-fixed quantities (such as the inflaton
perturbation) or even gauge-invariant variables such as the Bardeen
potentials --
in all cases the physical interpretation of the IR-divergences linked
to the metric/clocks of the local observer is the
same (a local dressing and rescaling of the metric or Hubble rate, see
section \ref{physicalproposal}) and
the prescription for the IR-safe correlator,
Eq.~(\ref{IRsafefirstway}), changes only in the 
fact that the chosen perturbation variables must be used in place of
$\zeta, \gamma$. Note that the action is indeed gauge
independent, and therefore the vertices are well-defined.

\vspace{0.4cm}

As a final comment, it can be useful to compare our proposal with the one in
Refs~\cite{Gerstenlauer:2011ti, Byrnes:2010yc}. As we have already
shown that their proposal is based on a quantity
$\kappa(q_0, L)^2 = e^{-2\zeta_{ir}}(e^{-\gamma_{ir}})_{ij}k^i k^j$,
see Eq.~(\ref{kappacut}), defined using $q_0, L$ to
define the classical fields $\zeta_{ir}, \gamma_{ir}$ via
the IR or large scale limit given by Eq.~(\ref{infraredfieldstasetal}).
We do not use any large scale limit in our case, but employ the 
quantum expectation 
value of the full quantum fields in our
Eqs.~(\ref{classicaleffectivethreemetric}), (\ref{replacement}). 

Moreover,  the IR-safe two-point correlator~\footnote{Actually,
  \cite{Gerstenlauer:2011ti, Byrnes:2010yc} discuss the spectrum,
  which is the two-point function rescaled by $k^3$, but this is not
  relevant here.} is also defined in a different way than in our case. 
Refs.~\cite{Gerstenlauer:2011ti, Byrnes:2010yc} define it via a
transformation $k \to \kappa(q_à, L)$ of the Fourier Transform of a
{\em spatial} average at {\em pair of points}:
 \beq \label{Akappacut}
  \mathcal{A}'(\kappa(q_0, L), \eta)_{\text{tree}}.
 \eeq
In other words, their proposal is to fix a certain physical cutoff
$q_0$ corresponding to the typical scale of the observation 
( $q_0^{-1}\sim |\vec x_2 -\vec x_1|$, with $\vec x_2, \vec x_1$ the
pair of points of the average),  
and define as ``infrared'' all wave numbers $k < q_0$, essentially
decomposing the fields as $\zeta = \zeta_{ir}^{k < q_0} + \zeta^{k > q_0}$,
promoting only $\zeta^{k > q_0}$ to quantum fields (analogously is done
for $\gamma$) and declaring the infrared safe correlators to be just
the tree-level result written in the new variables, see Eq.~(\ref{kappacut}). 

The first question that arises here is: since the correlator in 
Eq.~(\ref{Akappacut}) is 
only the tree-level contribution, supplemented with the IR effects in
Eq.~(\ref{kappacut}), what would be the IR-safe loop 
corrections to it in the proposal of \cite{Gerstenlauer:2011ti,
  Byrnes:2010yc}? The paper does not explicitly discuss this issue. It is however clear
that because of the decomposition 
$\zeta = \zeta_{ir}^{k < q_0} + \zeta^{k > q_0}$, the loop corrections would
have the form
 \beq
  \mathcal{A}'(\kappa, \eta)_{\text{loop, with IR cutoff $q_0$}}.
 \eeq
and thus the total correlator is:
 \beq \label{totcorrkappacut}
  \mathcal{A}'(\kappa, \eta)_{\text{tree}} + 
   \mathcal{A}'(\kappa, \eta)_{\text{loop, with IR cutoff $q_0$}}.
 \eeq 
It is straightforward to see then, that, although the ``IR-fields'' are
resummed in $\kappa$, equation (\ref{kappacut}) is indeed just given
by a coordinate transformation~\footnote{In fact, equation (\ref{Akappacut})
is similar to (\ref{expandamplitudek}), that is different from our
(\ref{IRsafefirstway}).}, and the final outcome amounts in effect
just to the use of an 
infrared cutoff $q_0$ in the loops. That is, there is in effect no
difference with the 
usual ``box cutoff'' approach, with all the issues that
we have discussed before.

It follows that there are big differences with our current
proposal: we {\em do not} introduce any infrared cutoff whatsoever,
not even an ``observational'' one like $q_0$, nor we decompose the
fields as $\zeta = \zeta_{ir}^{k < q_0} + \zeta^{k > q_0}$. What we do is really
to define a new perturbation theory on the basis of the quantum expectation
value $\overline h =\langle e^{2\widehat\zeta}e^{\widehat\gamma}
\rangle$ of the full quantum operator to define scales, instead than 
the standard perturbation theory around the background metric. The new
perturbative correlators are given by the series
Eq.~(\ref{IRsafefirstway}), and the higher order corrections to
correlators are non-trivial and do not reduce to the use of a cutoff
in the old perturbation theory. It is straightforward to generalize to all
$N$-point functions for both $\zeta$ and $\gamma$.


\subsection{Example at one-loop}\label{oneloopexample}

We now give a detailed example at one-loop  that our procedure do
cancel all IR divergences 
(Log-cutoff enhancements) while redefining the perturbative series, in
the theory with the Lagrangian in 
(\ref{actionsigma}).
To keep the example simple, let us consider a pure de Sitter case, so that 
$\zeta = 0, \sigma = \sigma(\vec x, t)$. In this case, the only
physical metric perturbations are gravitons. We wish to compute the
IR-safe two-point correlator
$\langle \sigma(\overline k, \eta) \sigma(\overline k', \eta)\rangle$,
at one loop, following our recipe culminating in equations
(\ref{IRsafefirstway}).  
 
We start by computing the one-loop correlator using the standard in-in
perturbation theory. We recall here the known results from Ref.~\cite{Giddings:2010nc}.
The one-loop in-in loop corrections are given by the diagrams
in Figs.~\ref{fig1} and \ref{fig2}. The relevant interaction
Lagrangians are obtained by expanding the full Lagrangian
in perturbation up to second order, and they read
 \beq\label{lthree}
  {\cal L}_3 = {a \ov 2}\gamma^{ij}\partial_i\sigma\partial_j\sigma\ ,
 \eeq
 \beq\label{lfour}
  {\cal L}_4 = -{a \ov 4}\gamma^{il}\gamma_{l}^{\;j}\partial_i\sigma\partial_j\sigma \, ,
 \eeq
where $a$ is scale factor of the de Sitter background. The computation
is performed by using Eq.~(\ref{ininamplitudes}), 
and the details are quite complicated and of little interest here. 
We will give some of the intermediate steps, details can be found in 
Refs.~\cite{Dimastrogiovanni:2008af, Giddings:2010nc}.
From the diagram in Fig.~\ref{fig1}, we obtain two contributions
 \be
  A_k(\eta) & = -8{\rm Re}\int {d^3q \ov (2\pi)^3}
  \int_{-\infty}^\eta{d\eta_1 \ov (H\eta_1)^2}
  \int_{-\infty}^{\eta_1}{d\eta_2 \ov (H\eta_2)^2}\omega_{ij,kl}(\vec q) k^ik^jk^kk^l
  W_q(\eta_1,\eta_2) W_{k-q}(\eta_1,\eta_2) \nonumber \\
  & \qquad\qquad\qquad\qquad\qquad\qquad\qquad\qquad\qquad\qquad \times W_k(\eta,\eta_1)W_k(\eta,\eta_2) \\
  B_k(\eta)& = 4{\rm Re}\int {d^3q \ov (2\pi)^3} \int_{-\infty}^\eta\prod_{i=1}^2
  {d\eta_i \ov (H\eta_i)^2}\omega_{ij,kl}(\vec q) k^ik^jk^kk^l
  W_q(\eta_1,\eta_2) W_{k-q}(\eta_1,\eta_2)
  W_k(\eta_1,\eta)W_k(\eta,\eta_2) ,
 \ee
where $W_a(\eta, \eta')$ is given by Eq.~(\ref{GenWhightman}). From Fig.~\ref{fig2}, we obtain
\beq
 C_k(\eta)= {\rm Re}\left[ (-2i) \int {d^3q \ov (2\pi)^3} \int
   \frac{d\eta'}{(H\eta')^2} \omega^{\phantom{i}l}_{i\phantom{l},lj} k^i k^j W_q(\eta',\eta'),
   W_k(\eta,\eta')^2\right]\ .  
\eeq 
The definition of the sum over graviton polarizations $\omega_{ij,kl}$ is
 \be \label{polarsum}
 \omega_{ij,kl}(\vec q) & =\sum_s \epsilon^s_{ij}(\vec q)\epsilon^{s*}_{kl}(\vec q)&& = 
  \delta_{ik}\delta_{jl}+ \delta_{il}\delta_{jk}-\delta_{ij}\delta_{kl} \\
  &&& ~~ + \delta_{ij}\vec {\hat q}_k\vec {\hat q}_l + \delta_{kl}\vec {\hat q}_i\vec {\hat q}_j- \delta_{ik}\vec {\hat q}_j\vec {\hat q}_l - \delta_{il}\vec {\hat q}_j\vec {\hat q}_k-\delta_{jk}\vec {\hat q}_i\vec {\hat q}_l-\delta_{jl}\vec {\hat q}_i\vec {\hat q}_k + \vec {\hat q}_i\vec {\hat q}_j\vec {\hat q}_k\vec {\hat q}_l\nonumber
\ee
where $\vec {\hat q}$ is the unit vector in the direction $\vec{q}$,
and $\epsilon_{ij}$ is the graviton polarization tensor.  After evaluating the integrals and regularizing it with IR and
ultraviolet cutoffs, we get
 \be
   \langle \sigma(k, \eta) \sigma(k', \eta)\rangle_{\text{one-loop}} &
   = A_k(\eta)+B_k(\eta)+C_k(\eta) \nonumber \\
   & = {2 \ov 15}\log(\Lambda_{IR}){H^4 \ov M_{\text{P}}^4}
    {1\ov (2\pi)^2 k^3}k^2\eta^2 \nonumber \\
   & ~~ + \left\{\text{finite terms and UV
      corrections}\right\}+{\cal O}\left({H^4 \ov M_{\text{P}}^4}\right)
 \ee
Therefore, adding the three-level result
 \beq
  \langle \sigma(k, \eta) \sigma(k', \eta)\rangle_{\text{tree}} =
  (2 \pi^3)\delta(k+k') \; {H^2 \ov M_{\text{P}}^2}{1 \ov 2k^3}(1+ k^2\eta^2) \, ,
 \eeq
the total result for the two-point correlator is~\footnote{In \cite{Giddings:2010nc}, only the result for
  $k\eta \to 0$ were presented.}:
 \be \label{ininoneloopsigmatwopoint}
  \mathcal{A}(k, \eta) & = \langle \sigma(k, \eta) \sigma(k, \eta)\rangle_{\text{tree}}
   + \langle \sigma(k, \eta) \sigma(k, \eta)\rangle_{\text{one-loop}}  & \nonumber \\
   & = \left[{H^2 \ov M_{\text{P}}^2}{1 \ov 2k^3}(1+ k^2\eta^2)+{2 \ov 15}\log(\Lambda_{IR}){H^4 \ov M_{\text{P}}^4}
    {1\ov (2\pi)^2 k^3}k^2\eta^2 \right.  \nonumber \\
   & ~~ \left. + \left\{\text{finite terms and UV
      corrections}\right\}\right]+ {\cal O}\left({H^4 \ov M_{\text{P}}^4}\right)
 \ee
We have reinstated the Planck mass to show explicitly the suppression
factors ${H^2 / M_{\text{P}}^2}$. Note that $|\gamma|^2 \sim {H^2 / M_{\text{P}}^2}$.

We now wish to apply our procedure to obtain the
IR-safe correlator up to one-loop order, starting from this
result. We follow the steps in 
section \ref{amendingprocedure}. The first one is to
compute the terms of the series in Eq.~(\ref{IRsafefirstway}) up to the order
of interest.  Since we are working up 
to one-loop, this means up to the order $\sim {H^4 \ov
  M_{\text{P}}^4}$, which is quadratic in $\gamma$. Looking at
Eq.~(\ref{ininoneloopsigmatwopoint}), we see that we need to compute up to
the second order derivative of $\langle \sigma(k, \eta) 
\sigma(k, \eta)\rangle_{\text{tree}}$, and no derivatives of
$\langle \sigma(k, \eta) \sigma(k, \eta)\rangle_{\text{one-loop}}$. 
The second order expansion in $\gamma$ gives
 \be
  {1 \ov 2}\langle\widehat\gamma^{ij}\widehat\gamma^{lf}\rangle
    \partial_{\gamma^{ij}_{_{\bar h}}}\partial_{\gamma^{lf}_{_{\bar h}}}
     \langle \sigma(k, \eta)\sigma(k', \eta)\rangle_{\text{tree}}\big|_{k^2 = \overline k^2}
   & = {1 \ov 2}\sum_{i, j} \langle\widehat\gamma^{il}\widehat\gamma_l^{\;j}\rangle \overline k_i \overline k_j
   \partial_{k^2}\langle \sigma(k, \eta)\sigma(k',\eta)\rangle_{\text{tree}}\big|_{k^2 = \overline k^2}
   \nonumber \\
   & + {1 \ov 2}\sum_{i, j}\langle \widehat\gamma^{ij}\widehat\gamma^{kl}\rangle 
   \overline k_i \overline k_j \overline k_k \overline k_l \partial_{k^2}^2
  \langle \sigma(k, \eta)\sigma(k', \eta)\rangle_{\text{tree}}\big|_{k^2 = \overline k^2}.
 \ee
After some manipulation, we obtain
the contributions
 \begin{multline}
   \label{twokeystwopointderivative}
  {1 \ov 2}\sum_{i, j} \langle\gamma^{il}\gamma_l^{\;j}\rangle \overline k_i \overline k_j
   \partial_{k^2}\langle \sigma(k, \eta) \sigma(k', \eta)\rangle_{\text{tree}}\big|_{k^2 = \overline k^2}  =  
    -{2 \ov 3}\log(\Lambda_{IR}){H^4 \ov M_{\text{P}}^4}
     {1\ov (2\pi)^2 \overline k^3}\overline k^2\eta^2 + \\
    ~~ + \left\{\text{disconnected finite terms and UV
      corrections}\right\}
 \end{multline}
 \begin{multline}
    \label{fourkeystwopointderivative}
  {1 \ov 2}\sum_{i, j}\langle \gamma^{ij}\gamma^{kl}\rangle 
   \overline k_i \overline k_j \overline k_k \overline k_l \partial_{k^2}^2
  \langle \sigma(k, \eta)\sigma(k', \eta)\rangle_{\text{tree}}\big|_{k^2 = \overline k^2}  =
    {4 \ov 5}\log(\Lambda_{IR}){H^4 \ov M_{\text{P}}^4}
     {1\ov (2\pi)^2 \overline k^3}\overline k^2\eta^2 + \\
    ~~ + \left\{\text{other disconnected finite terms and UV
      corrections}\right\}.
 \end{multline}
We now apply Eqs.~(\ref{IRsafefirstway}), subtracting
Eqs.~(\ref{twokeystwopointderivative},~\ref{fourkeystwopointderivative})
form Eq.~(\ref{ininoneloopsigmatwopoint}), to obtain 
 \be
  \mathcal{A}_{\text{IR-safe}}(\overline k, \overline k', \eta) & = (2 \pi^3)\delta(\overline k+\overline k') &
  & \left[{H^2 \ov M_{\text{P}}^2}{1 \ov 2\overline k^3}(1+ \overline k^2\eta^2)+
  \right.  \nonumber \\
   & & & \left. + \left\{\text{new finite terms and UV
      corrections}\right\}\right]+ {\cal O}\left({H^4 \ov M_{\text{P}}^4}\right).
 \ee
As  we can see, the dependence of
$\log(\Lambda_{IR})$ has disappeared and there is no trace of any
other IR regulator. The new finite terms and ultraviolet corrections
are given by the difference between those in equation 
(\ref{ininoneloopsigmatwopoint}) and the disconnected contributions in
(\ref{twokeystwopointderivative}), (\ref{fourkeystwopointderivative});
they are non-zero as (\ref{ininoneloopsigmatwopoint}) is not given by
disconnected diagrams.

\subsection{$N$-point functions and non-Gaussianities}\label{Npointfunctions}

What we have said about the two-point function can be extended to the
most general case of $N$-point functions. 
It is interesting then to discuss the loop correction to the
three-point function and the bispectrum, which is the first indication of
non-Gaussianities in the primordial perturbations. 

The latter ones are usually
quantified with a set of parameters. The one which is related to the
three-point function is called $f_{NL}$, having defined
 \beq\label{shape-func}
  \langle
  \zeta_{\vk_1}(\eta)\zeta_{\vk_2}(\eta)\zeta_{\vk_3}(\eta)\rangle
    \equiv \delta(\sum_i \vk_i) (2\pi)^3 F(\vk_1,\vk_2,\vk_3, \eta),
 \eeq
and written the bispectrum as
 \beq
   F(\vk_1,\vk_2,\vk_3, \eta) \equiv -{6 \ov 5}
   f_{NL}[P^{(\zeta)}P^{(\zeta)} + \text{2 permutations}]
 \eeq 
The parameter $f_{NL}$ will receive corrections at one-loop. 
In particular, in the standard approach, it suffers from IR
Logarithmic-enhancements (although mitigated by some dependence on the
spectral index $1-n$) \cite{Giddings:2010nc}. Such corrections can 
make non-Gaussianities much more pronounced theoretically than what is
actually measured, see~\cite{WMAP}. 

Instead, if we use our IR-safe proposal to calculate the
three-point function corrected up to one loop, one does not find such
Logarithmic-enhancements, but only the ultraviolet corrections (which are
renormalized). We will not compute them here as it goes beyond
the scope of this paper.

\section{Conclusion}\label{conclusion}

In this work we have proposed a method to solve the IR issues of
cosmological correlators. The solution we propose here is based on the
physical principle -- every observable (correlator) should be defined in terms of a
    local measurement. Of course, when one in practice
      approximates the ensemble averaging of the correlators with the
      spatial averaging, one introduces by definition a certain degree of 
      non-locality. We have discussed
      the connection between the IR issue and this approximation
    at the end of section \ref{GeneralBehaviour}.

Our proposal consists in a modification of the formalism enabling us
to ask physical questions.
In particular, we emphasize that quantities must be defined locally by
local observers and we implement this principle by defining a new
perturbation theory in the in-in formalism.

We do not see any breaking of perturbation theory due to IR
corrections~\cite{ArkaniHamed:2007ky}, as
our definition of the perturbation theory is IR-safe. Furthermore,
there is no ambiguity related to the choice of regularization of the
IR divergent integrals, as the final result does not depend on it.

\section*{Acknowledgments}
The authors would like to thank David Lyth, Mischa Gerstenlauer, Arthur
Hebecker and Gianmassimo Tasinato for helpful discussions.
D.C. is supported by a Postdoctoral F.R.S.-F.N.R.S. research
fellowship via the Ulysses Incentive Grant for the Mobility in Science
(promoter at the Universit\'e de Mons: Per Sundell).



\end{document}